\documentclass[final, aps, reprint, lengthcheck, footinbib, preprintnumbers, showpacs, citeautoscript, superscriptaddress, balancelastpage]{revtex4-1}

\usepackage[T1]{fontenc}
\usepackage[english]{babel}
\usepackage{amsmath}
\usepackage{amssymb}
\usepackage{stix}
\usepackage{braket}
\usepackage{graphicx}
\usepackage[colorlinks, urlcolor=blue, citecolor=blue, linkcolor=blue, pdfstartview=FitH]{hyperref}
\usepackage[all]{hypcap}
\usepackage[dvipsnames]{xcolor}
\usepackage{subfigure}

\begin{document}

\def\affiEII{Experimentelle\ Physik\ 2, Technische\ Universit\"at\ Dortmund, D-44221 Dortmund, Germany}
\def\affiEIII{Experimentelle\ Physik\ 3, Technische\ Universit\"at\ Dortmund, D-44221 Dortmund, Germany}
\def\affiSOLAB{Spin\ Optics\ Laboratory, Saint~Petersburg\ State\ University, 198504 St.~Peterbsurg, Russia}
\def\affiIOFFE{Ioffe\ Institute, Russian\ Academy\ of\ Sciences, 194021 St.~Petersburg, Russia}

\title{Subsecond nuclear spin dynamics in \emph{n}-GaAs}

\author{P.~S.~Sokolov}
\affiliation{\affiEII}
\affiliation{\affiSOLAB}
\author{M.~Yu.~Petrov}
\author{K.~V.~Kavokin}
\author{M.~S.~Kuznetsova}
\author{S.~Yu.~Verbin}
\author{I.~Ya.~Gerlovin}
\affiliation{\affiSOLAB}
\author{D.~R.~Yakovlev}
\affiliation{\affiEII}
\affiliation{\affiIOFFE}
\author{M.~Bayer}
\affiliation{\affiEII}
\affiliation{\affiIOFFE}
\date{\today}

\begin{abstract}
We use time-resolved detection of the Hanle effect and polarized photoluminescence with dark intervals to investigate the buildup and decay of the spin polarization of nuclei interacting with donor-bound electrons in $n$-doped GaAs. Strong hyperfine coupling defines the millisecond timescale of the spin dynamics of these nuclei, as distinct from the nuclei far from impurity centers, characterized by a thousand times longer spin-relaxation time. The dynamics of spin polarization and relaxation attributed to the nuclei inside the donor orbit is observed on the time scale from 200 to 425 ms.
\end{abstract}

\maketitle

\section{\label{sec:intro}Introduction}

The spin physics of electrons and nuclei in semiconductors, after 50 years of development, continues to reveal bright phenomena, mainly in advanced nanostructures~\cite{DyakonovSpinBook, UrbaszekRMP13}.
However, even in otherwise very well studied bulk semiconductor materials, like GaAs, the understanding of the electron-nuclear spin dynamics, first discovered in the 1970s \cite{PagetPRB77} (see also Ref.~\cite{OO}) has considerably progressed during recent years~\cite{SokolovPRB17, KoturPRB16, KoturPRB18, VladimirovaPRB18, ScalbertPRB17}.
The application of modern optical techniques like Faraday rotation~\cite{GiriPRB12,GiriPRL13}, spin noise~\cite{RyzhovAPL15, RyzhovScRep16} and time-resolved photoluminescence~\cite{SokolovPRB17,KoturPRB18}, combined with more traditional techniques of magneto-optical spectroscopy and nuclear magnetic resonance (NMR)~\cite{SuterJMR04, SuterJMR14}, gave the opportunity to identify the basic mechanisms of spin and energy relaxation in the intertwined spin systems of nuclei and resident charge carriers~\cite{VladimirovaPRB17}.

The pattern of optical pumping of nuclear spins in GaAs at low temperatures was outlined by Paget in 1982~\cite{PagetPRB82}: optically oriented spins of localized electrons dynamically polarize nuclear spins in the vicinity of localization centers (typically, donor impurities); the nuclear spin polarization spreads over the crystal by spin diffusion mediated by the dipole-dipole interaction. Paget measured the time increment of the dynamic polarization by using optically detected NMR in high-purity GaAs and found it to be about $3$~s.

The~theory \cite{Abragam, DP_Nuclei_1973} suggested that the relaxation time of nuclear spins due to their hyperfine coupling with electrons, $T_{1e}$, is inversely proportional to the mean squared hyperfine frequency $\omega_\mathrm{hf}$ and to the electron spin correlation time $\tau_c$
\begin{equation}
T_{1e} \propto \left[ \braket{\omega^2_\mathrm{hf}} \tau_c \right]^{-1}.
\label{eq:T_1e}
\end{equation}
With the knowledge of the electron spin density in the probed region (at the distance of one Bohr radius of the hydrogen-like shallow donor from the impurity center), this allowed Paget to estimate the spin correlation time of the donor-bound electron, $\tau_c \approx 25$~ps.
This short time was attributed to spin exchange with itinerant photoexcited electrons in the conduction band. 
Theoretical calculations of the exchange scattering rates \cite{PagetPRB81, KavokinSST08} supported this interpretation.
Recently, a $T_{1e}$ of about $10$~s was measured, using off-resonant Faraday rotation \cite{GiriPRL13}, for nuclei interacting with localized electrons in a structure with an electron gas, where strong exchange scattering occurred even in the absence of optical pumping.
From that experiment, Giri\ \emph{et~al}.~\cite{GiriPRL13} estimated $\tau_c \approx 10$~ps, corroborating the Paget model~\cite{PagetPRB82}.

In intentionally $n$-doped semiconductors with donor concentrations $n_D$ below the metal-to-insulator transition, characterized by long electron-spin lifetimes \cite{DzhioevPRB02} and, therefore, prospective for achieving a high polarization of nuclear spins, the electron spin correlation times are longer.
For instance, in $n$-GaAs with $n_D \approx 10^{15}$ cm$^{-3}$ spin correlation time $\tau_c$ is about $300$~ps~\cite{DzhioevPRB02}, which suggests a very efficient nuclear-spin relaxation near donors, with  $T_{1e} \approx 300$~ms.
As a result, donor impurities act as ``killer centers'' for nuclear polarization; the nuclei situated far from the donors lose their spin polarization by spin diffusion to donors~\cite{KoturPRB16}, which determines a much longer time ($T_\mathrm{bulk} \sim 10^2$~s), characterizing their relaxation~\cite{Note1}.

While nuclear polarization and relaxation by spin diffusion from or to donors in $n$-type semiconductors are well documented experimentally~\cite{KalevichJETPLett82}, the millisecond time scale constants of dynamic polarization and hyperfine relaxation of those nuclei, which are directly coupled to donor-bound electrons, have not been experimentally measured, for the lack of appropriate techniques.
In this work, we use a method based on measuring the Hanle effect (depolarization of photoluminescence by magnetic fields) with millisecond time resolution~\cite{SokolovPRB17, KoturPRB18} to study the nuclear spin dynamics in the vicinity of donors in $n$-GaAs.
We measure both the rise time of the nuclear spin polarization under optical pumping and its decay time in the dark.
The relation between these times supports the suggestion of Ref.~\cite{PagetPRB82} on the important role of photoexcited electrons in the nuclear spin dynamics.

\section{\label{sec:experiment}Experimental Details}

\begin{figure*}[t!]
\subfigure{\label{fig:PLHanle:a}}
\subfigure{\label{fig:PLHanle:b}}
\subfigure{\label{fig:PLHanle:c}}
\includegraphics[width=\textwidth,clip]{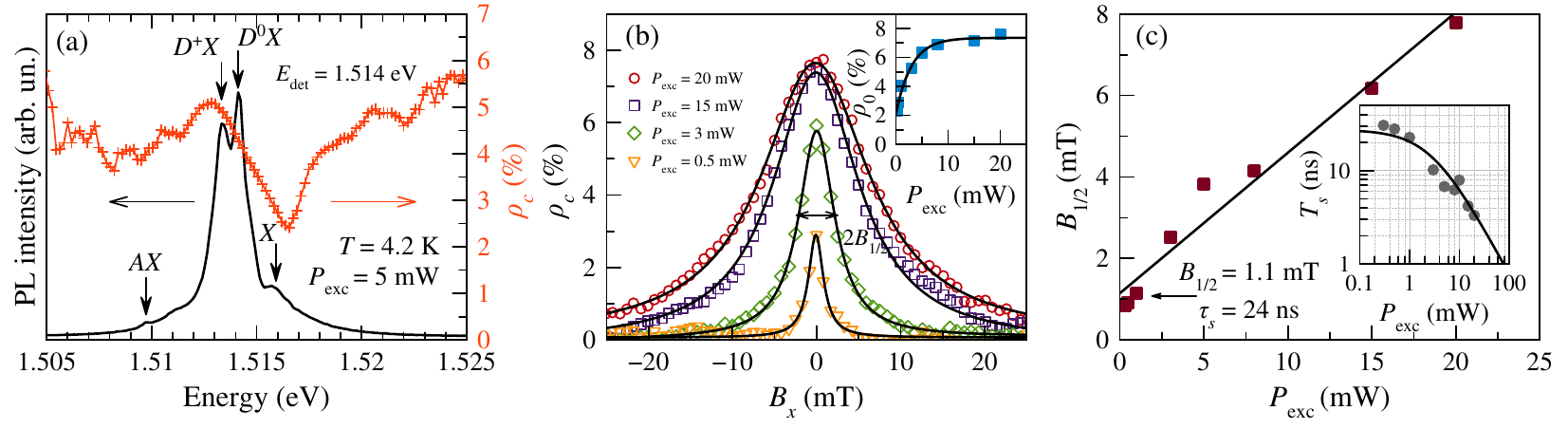}
\caption{(a) The PL spectrum of $n$-doped GaAs measured at $B=0$ (solid line) and the spectral dependence of the PL circular polarization degree (crosses). (b) The Hanle curves of $D^0X$ line measured for fast polarization modulation $f_\mathrm{mod} (\sigma^+/\sigma^-)=50$~kHz at various pumping powers (symbols) and their fitting with Lorentzians. Inset shows the power dependence of the Hanle curve amplitude $\rho_0$ (squares) and its fit with an exponential function (line). (c) Power dependence of $B_{1/2}$ (squares) and its linear fit (line). Inset shows the power dependence of the electron spin lifetime (circles) extracted from the HWHM of the Hanle curves and its fit with a rational function (line).}
\label{fig:PLHanle}
\end{figure*}

The studied sample is the GaAs epitaxial layer grown by liquid-phase epitaxy on top of a semi-insulating (001) GaAs substrate.
The $20$-$\mu$m epitaxial layer was doped by Si providing a donor concentration $n_D = 4 \times 10^{15}$~cm$^{-3}$.
The photoluminescence (PL) is excited by a tunable Ti:Sapphire laser operated at $E_\mathrm{exc}=1.55$~eV corresponding to the absorption edge of the GaAs band-to-band transition. The laser is focused on the sample surface through an achromatic doublet (focal distance $F=200$~mm) into a spot of about $80$~$\mu$m in diameter ($1/e^2$ width) corresponding to the excitation density of about $W_\mathrm{exc} \approx 40$ W/cm$^2$ for the excitation power $P_\mathrm{exc} = 1$~mW.
The sample is placed in a cryostat that provides the possibility of changing the sample temperature from $T=4.2$~K up to room temperature.
A set of Helmholtz coils allows us to apply magnetic fields along the optical axis (Faraday geometry, $B_z$) and in the perpendicular direction (Voigt geometry, $B_x$) simultaneously.
The helicity of optical excitation is controlled by an electro-optical modulator that produces a time-dependent phase shift between the linearly polarized components of the light wave.
This allows us to implement excitation protocols with a fast switch between circular right ($\sigma^+$) and circular left ($\sigma^-$) light polarizations.
The PL is collected and spectrally filtered by a $0.5$-m spectrometer followed by a gated single-photon counter (SPC).

The experimental setup has been extended in order to realize various temporal excitation-detection protocols. The circular polarization of the excitation could be modulated between $\sigma^+$ and $\sigma^-$ with frequency $f_\mathrm{mod}$ from sub-Hz up to $50$~kHz.
The PL counting is performed in a time window ranging from $20$~$\mu$s to $500$~ms delayed from the excitation trigger by a varied time delay $\Delta t$.
The analysis of the PL circular polarization degree is done by using a photoelastic modulator (PEM) followed by a large-aperture Glan-Taylor polarizer.
The intensities of the $\sigma^+$ and $\sigma^-$ PL polarization components are obtained by gating the pulses generated by the SPC synchronously with the PEM modulation and accumulating the intensities $I_\mathrm{co}$ and $I_\mathrm{cross}$ corresponded to co-circular and cross-circular PL polarization helicities with respect to the excitation in two separate channels of the two-channel photon counting device (Stanford Research SR400).
The degree of circular polarization is obtained as $\rho_c = (I_\mathrm{co} - I_\mathrm{cross}) / (I_\mathrm{co} + I_\mathrm{cross})$ with the accuracy of about $0.1$\% determined by a number of photon counts accumulated during a single measure. 
The excitation modulation, the PL gating time window, and the photon counting gates are synchronized by a precise digital delay generator (Quantum Composer 9520). 
Thereby, various problems could be addressed; particularly the time-dependent measurement of $\rho_c$ in the Hanle effect with a polarization modulation is performed~\cite{SokolovPRB17}.

To eliminate the possible impact of the nuclear spin polarization, the helicity of the pumping light ($\sigma^+ / \sigma^-$) can be modulated at a high frequency, $f_\mathrm{mod}$, exceeding several tens of kHz~\cite{OO}.
To study the nuclear spin dynamics, $f_\mathrm{mod}$ is varied in a range from several mHz to hundreds of Hz.

\section{\label{sec:experiment}Results and Discussion}

Four distinct peaks corresponding to recombination of the free exciton ($X$), the exciton bound to neutral and charged donors ($D^0X$ and $D^+X$), as well as the exciton-acceptor complex ($AX$) can be identified in the PL spectrum shown in Fig.~\ref{fig:PLHanle:a}.
The figure also demonstrates a nonmonotonic spectral behavior of $\rho_c$.
Following Ref.~\cite{SokolovPRB17}, part of the spectrum corresponding to the $D^0X$ transition is selected to measure the spin polarization of electrons interacting with the nuclear spins situated in the vicinity of donor centers.

In the case when the frequency of polarization modulation of light between $\sigma^+$ and  $\sigma^-$ $f_\mathrm{mod}$ exceeds $10$~kHz, the nuclear spin system remains unpolarized and the dependence $\rho_c(B)$ (the Hanle curve) obeys Lorentzian law with the half width at half maximum (HWHM), $B_{1/2}$~\cite{OO}.
In Fig.~\ref{fig:PLHanle:b}, a set of Hanle curves measured at different powers of excitation, using fast modulation of excitation at $f_\mathrm{mod} = 50$~kHz, is shown.
Since at high enough modulation frequency $f_\mathrm{mod} \gg T_2^{-1}$, where $T_2$ is the thermal equilibrium establishment time, than nuclear spin polarization is negligible~\cite{OO}. This condition should  be fulfilled for all different isotopes because of different neighboring spins and, therefore, different dipole-dipole or quadrupole interaction exhibited by these nuclei~\cite{ChenPRB11}.

The Hanle curves shown in Fig.~\ref{fig:PLHanle:b} are, to a good approximation, Lorentzians, $\rho_c(B_x)=\rho_0 / \left( 1 + B_x^2/B_{1/2}^2 \right)$.
This allows us to determine the electron spin lifetime, $T_s = \hbar / (\mu_B g_e B_{1/2})$, where $g_e$ is the electron $g$-factor in GaAs~\cite{OO}.
As seen from the inset in Fig.~\ref{fig:PLHanle:c}, $T_s$ becomes shorter with increasing the excitation power, due to the increase of recombination rate of electrons with photoexcited holes and, possibly, to the effect of hot carriers on spin relaxation.
By plotting the power dependence of the $B_{1/2}$ values and taking the cut-off at zero pump power with $B_{1/2}=1.1$~mT and $|g_e| = 0.44$, the intrinsic electron-spin relaxation time $\tau_s=24$~ns is extracted [see Fig.~\ref{fig:PLHanle:c}].
The inset in Fig.~\ref{fig:PLHanle:b} shows the power dependence of the electron spin polarization degree determined at zero magnetic field, $\rho_0$.
At pumping powers $P_\mathrm{exc} \gtrsim 5$~mW,  $\rho_0$ is saturated, while the spin lifetime continues to fall. Since the excitation power of  $P_\mathrm{exc}=5$~mW provides efficient pumping of the nuclear spin system, we use it in all the experiments described in the following text.

As discussed in the Introduction, the nuclear spin dynamics in $n$-GaAs is characterized by, at least, two times cales: the short time of the hyperfine relaxation inside the donor orbit (\(\sim\)$10^{-1}$~s) and the long time of spin diffusion far from donor centers (\(\sim\)$10^2$~s)~\cite{KalevichIzv82}.
Both times are observed in our experiments. 

\begin{figure}[t]
\subfigure{\label{fig:Demagnetization:a}}
\subfigure{\label{fig:Demagnetization:b}}
\subfigure{\label{fig:Demagnetization:c}}
\includegraphics[width=\columnwidth,clip]{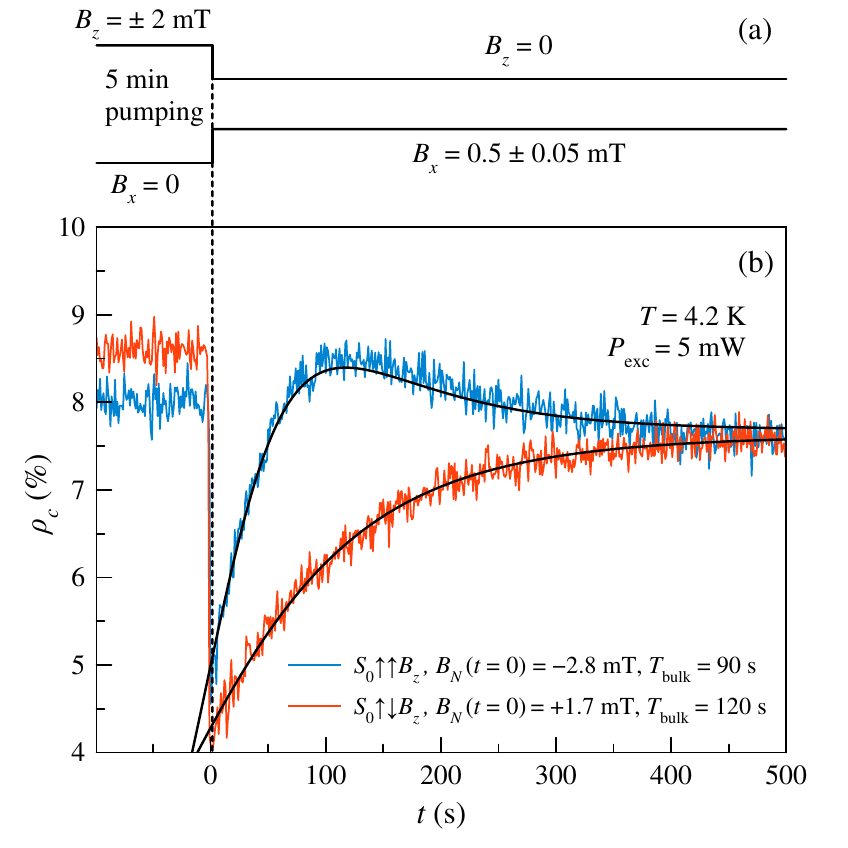}
\caption{Time dependence of the PL circular polarization degree after demagnetization from the longitudinal $B_z = + 2$~mT (blue curve) and $B_z = -2$~mT (red curve) fields in a small transverse magnetic field $B_x = 0.5 \pm 0.05$~mT and its fitting (solid lines) with Eq.~\eqref{eq:long_nuc_field}.
}\label{fig:Demagnetization}
\end{figure}

To evaluate the spin relaxation time of the bulk nuclei we used the experimental protocol of Ref.~\cite{KalevichJETPLett82}, shown in Fig.~\ref{fig:Demagnetization:a}.
The spin system is pumped with circularly polarized light during $5$~min in the longitudinal magnetic fields $B_z = \pm 2$~mT.
After that (at $t = 0$~s), the longitudinal magnetic field is switched off and a small transverse field  $B_x \approx 0.5 \pm 0.05$~mT is switched on.
With switching off both the pumping light and the longitudinal field, the nuclear spin cooling stops and the further dynamics of the electron-spin polarization is determined by nuclear demagnetization in the small transverse magnetic field $B_x$.
As a result of this demagnetization, the degree of electron-spin polarization approaches the level corresponding to the stationary value $S_0$ in the field $B_x$  [see  Fig.~\ref{fig:Demagnetization:b}].
Depending on the mutual orientation of the mean electron spin and $B_z$ during the cooling period, the Ovehauser field, develops in the direction along $B_z$ or opposite to $B_z$. 
Correspondingly, at $t=0$~s the $B_N$ is directed either along $B_x$ or opposite to it~\cite{OO}.
As a result, the curves shown in Fig.~\ref{fig:Demagnetization:b} have different shapes.

In order to relate the time dependence of $\rho_c$, shown in Fig.~\ref{fig:Demagnetization:b}, to the Overhauser field $B_N$ exponential decay, the model represented in Ref.~\cite{GiriPRB12} can be used. 
At $t>0$, $\rho_c (t)$ represents the time evolution of a point in the Hanle curve where the total magnetic field, including the time-dependent Overhauser field, acting on the electron spin is aligned along or against the applied external magnetic field $B_x$. 
It reads
\begin{equation}
\rho_c \propto \left[  1 + \left( B_x + B_{N}(t=0) e^{-t/T_\mathrm{bulk}} \right)^2 / B_{1/2}^2 \right]^{-1},
\label{eq:long_nuc_field}
\end{equation}
where $B_{1/2}$ is the Hanle curve HWHM measured at the same excitation power with rapidly modulated pumping helicity [see Fig.~\ref{fig:PLHanle:b}].
Here, $B_N(t=0)$ and $T_\mathrm{bulk}$ are obtained from fitting.
The dynamics of the Overhauser field decay we characterized by a monoexponential process with characteristic time $T_\mathrm{bulk}$.
We obtained $T_\mathrm{bulk}=90$~s for $B_z = +2$~mT and $T_\mathrm{bulk}=120$~s for $B_z = -2$~mT.
These times are of the same order of magnitude as those obtained earlier for the nuclear spin relaxation in bulk $n$-GaAs~\cite{VladimirovaPRB17}.
The difference between these relaxation times is due to a weak dynamic polarization of nuclear spins by the circularly polarized pump light in the transverse magnetic field.
This contribution to the nuclear polarization occurs due to nuclear spin cooling in the effective electron field, known as the Knight field.
It creates the Overhauser field that is always directed along $B_x$~\cite{OO}. Being superimposed on the relaxation of the nuclear polarization, this effect either increases or decreases $T_\mathrm{bulk}$, depending on the initial direction of the Overhauser field.

The time resolution used in the experiment represented in Fig.~\ref{fig:Demagnetization} is insufficient to detect the fast millisecond dynamics of the nuclei in the vicinity of the donor center.
In order to observe such dynamics, we use a rapid alternation of polarization and gating the light intensity.
This allows us to measure the dynamics of the onset of the nuclear spin polarization under pumping, as well as of its decay in the dark.

\begin{figure}[t]
\subfigure{\label{fig:Repump:a}}
\subfigure{\label{fig:Repump:b}}
\subfigure{\label{fig:Repump:c}}
\subfigure{\label{fig:Repump:d}}
\includegraphics[width=\columnwidth,clip]{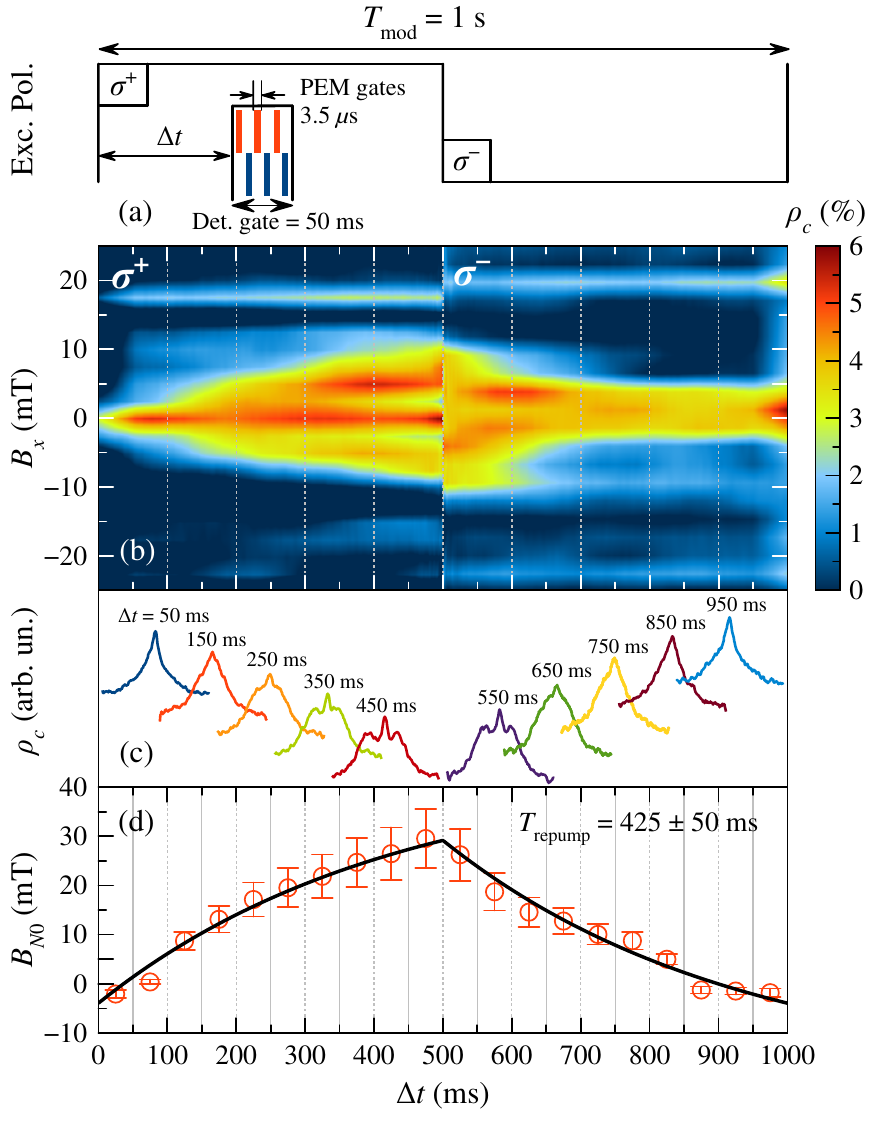}
\caption{(a) Schematics of the measurement protocol. (b) Dynamics of $\rho_c$ measured for the $\sigma^+$ and $\sigma^-$ half periods ($T_\mathrm{mod} = 1$~s) at different $B_x$ and fixed $B_z=+2$~mT. (c) Hanle curves corresponding to the gated detection in the time intervals shown in panel (b). (d) Magnitude of the Overhauser field, $B_{N0}$, extracted from fitting the Hanle curves (symbols) with Eq.~\eqref{eq:BNt} (solid line).
}\label{fig:Repump}
\end{figure}

\begin{figure}[b]
\subfigure{\label{fig:Model:a}}
\subfigure{\label{fig:Model:b}}
\includegraphics[width=\columnwidth,clip]{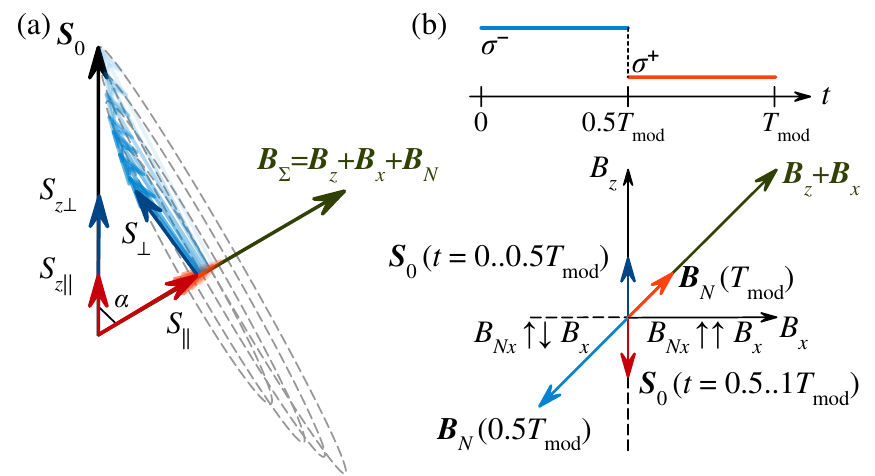}
\caption{Diagram showing the various components of the electron~(a) and nuclear~(b) spins in an tilted external magnetic field (the Hanle effect for oblique geometry). The sign of the nuclear field is dependent on the orientation of the pump ($\sigma^+$ or $\sigma^-$) as follows from the inset of panel (b).
}\label{fig:Model}
\end{figure}

To demonstrate the presence of a fast nuclear spin pumping dynamics, we used the time protocol shown in Fig.~\ref{fig:Repump:a}, where the pumping helicity is alternated each $500$~ms.
The extracted time dependence of $\rho_c$ is mapped in Fig.~\ref{fig:Repump:b} for external fields $B_x$ varied from $-25$ to $+25$~mT and the fixed value of $B_z = +2$~mT.
The time resolution in Fig.~\ref{fig:Repump:b} is $\delta t = 2$~ms. $B_x$ is scanned very slowly because such a measurement requires a long accumulation time for the time traces of the $\rho_c$ dynamics.
Each slice of Fig.~\ref{fig:Repump:b} at a given time interval represents a Hanle curve.
During the first $100$~ms, the Hanle curve is almost Lorentzian, with the maximum at $B_x = 0$.
One can see the development of additional peaks, which are shifting towards higher fields with time.
Such a dynamics occurs due to the onset of the Overhauser field compensating the external magnetic field.
In the next half-period, the pumping helicity is changed for the opposite and the Overhauser field adds to the external field, which results in the disappearance of side peaks and narrowing of the Hanle curve by the end of the modulation period.

To examine further details of the experimental data shown in Fig.~\ref{fig:Repump:b}, we use the time-resolved Hanle measurement.
With the same period $T_\mathrm{mod}=1$~s of excitation polarization, the $50$-ms-long detection gate gradually scans  over the modulation period [Fig.~\ref{fig:Repump:a}].
The Hanle curves are measured by scanning the $B_x$ field [with the scanning rate similar to that used for the measurements in Fig.~\ref{fig:Repump:b}] at fixed $B_z = +2$~mT for various delay times $\Delta t$.
Examples of the Hanle curves obtained using this detection protocol are shown in Fig.~\ref{fig:Repump:c}.
With this method, we qualitatively recover the time evolution of the Hanle curve shown in the colormaps of Fig.~\ref{fig:Repump:b}.
However, at the expense of lower time resolution, the signal-to-noise ratio is increased, allowing us to quantitatively analyze the evolution of the Overhauser field.

The appearance of the side maxima in the Hanle curves can be understood as a consequence of the partial compensation of the external field by the Overhauser field $B_N$.
The electron-spin polarization is governed by the balance of the generation of oriented electrons by circularly polarized light $S_0$ and their spin depolarization caused by Larmor precession in the total magnetic field $B_\Sigma=B+B_N$, as well as by the electron-spin relaxation.
In zero magnetic field, $S_z(0) = S_0/(1+\tau/\tau_s)$, where $\tau$ is the electron lifetime and the dependence of the $z$~projection of the electron spin on the magnetic field (the Hanle curve) is given by the following expression [see Fig.~\ref{fig:Model:a}]:
\begin{equation}
\begin{split}
&S_z(B) = S_{z\parallel} + S_{z\perp} =\\
&S_z(0) \frac{(B_z + B_{Nz})^2}{B_\Sigma^2} + S_z(0) \frac{(B_x + B_{Nx})^2}{B_\Sigma^2} \frac{1}{1+B_\Sigma^2/B_{1/2}^2},
\end{split}
\label{eq:Sz}
\end{equation}
where $B_\Sigma^2 = (B_z + B_{Nz})^2 + (B_x + B_{Nx})^2$ is the total field acting on the electron spin and $B_{1/2}$ is the Hanle amplitude and HWHM in the absence of the nuclear spin polarization, while $B_{Nx}$ and $B_{N_z}$ are the Overhauser field projections.
The time dependence of $B_{Nx}$ and $B_{N_z}$ can be evaluated [see Fig.~\ref{fig:Model:b}] as
\begin{subequations}
\begin{align}
	B_{Nx} = \frac{B_z B_x}{B_z^2 + B_x^2} B_{N0}(t),\label{eq:BNx}\\
	B_{Nz} = \frac{B_z^2}{B_z^2 + B_x^2} B_{N0}(t).\label{eq:BNz}
\end{align}
\end{subequations}
Here, $B_{N0}(t)$ is a scalar function of time representing the magnitude of the Overhauser field obtained under repumping conditions with a step-like switching of $S_0$ from $+|{S_0}|$ to $-|{S_0}|$.
As follows from Eq.~\eqref{eq:BNz}, $B_{N0}$ gives the value of the Overhauser field at $B_x = 0$.
The temporal dependence of $B_{N0}$ within a half-period is determined by the convolution of the step-like and the exponential decay functions
\begin{equation}
B_{N0}(t) = \mathcal{K} S_0 \left( 1 - \frac{2 e^{-t/T_\mathrm{repump}}}{1+e^{-T_\mathrm{mod}/(2T_\mathrm{repump})}} \right) + \mathcal{P} S_0^2,	\label{eq:BNt}
\end{equation}
where $\mathcal{K}$, $\mathcal{P}$, and $T_\mathrm{repump}$ are the fitting parameters.

Equation~\eqref{eq:Sz} is used to fit the Hanle curves shown in Fig.~\ref{fig:Repump:c} with the following set of the fixed parameters: $S_z(0) = 0.06$, $B_z = +2$~mT and $B_{1/2} = 4$~mT, and a varied parameter $B_{N0}(t)$.
From this fit, $B_{N0}(t)$ is obtained, which dynamics is shown by the symbols in Fig.~\ref{fig:Repump:d}.
Fitting this dependence by Eq.~\eqref{eq:BNt} gives a characteristic time $T_\mathrm{repump} = 425 \pm 50$~ms.
The coefficient $\mathcal{K} S_0 = 31 \pm 3$~mT reflects the efficiency of the dynamic polarization of the nuclear spins inside the donor orbit.
The time-independent contribution to $B_{N0}$ given by $\mathcal{P} S_0^2 = 13 \pm 2$~mT results from the nuclear spin cooling in the oscillating Knight field~\cite{SokolovPRB17, OO}.

\begin{figure}[t]
\subfigure{\label{fig:FastRelaxation:a}}
\subfigure{\label{fig:FastRelaxation:b}}
\includegraphics[width=\columnwidth,clip]{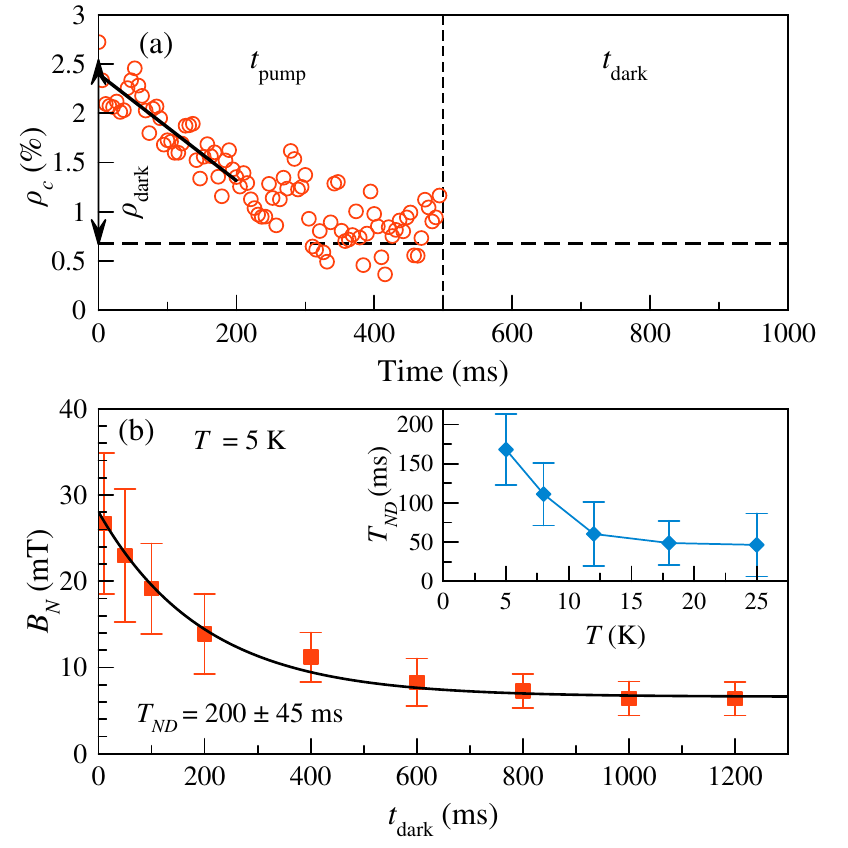}
\caption{(a) Time dependence of the $\rho_c$ for the presence of circularly polarized pump. (b) The Overhauser field values calculated for different dark time intervals using Eq.~\eqref{eq:nuc_field} (squares) and its fitting with an exponential decay function (solid line). Inset shows the temperature dependence of the nuclear spin-relaxation time.
}\label{fig:FastRelaxation}
\end{figure}

To study the nuclear spin relaxation mediated by the hyperfine interaction with donor bound electrons in the absence of illumination, we use the method described in detail in Ref.~\cite{KoturPRB18}.
The laser pulses are cut out from the continuous wave laser beam with an acousto-optical modulator controlled by a digital pulse generator.
During the dark and bright time intervals, the external magnetic fields $B_{z}=+2$~mT and $B_{x}= + 5$~mT are applied.
The PL kinetics is analyzed using a multi-channel photon-counting system.
We use the two-stage protocol [see Fig.~\ref{fig:FastRelaxation:a}] having two alternating time intervals: the bright interval $t_\mathrm{pump} = 500$~ms is followed by a dark time of various duration (from $t_\mathrm{dark} = 10$~ms to $t_\mathrm{dark} = 1.3$~s).

During the bright interval, the Overhauser field is induced by pumping with circularly ($\sigma^+$) polarized light and the time-resolved PL measurement in both circular polarizations gives the electron-spin-polarization dynamics under the development of the Overhauser field.
During the dark time, the Overhauser field relaxes.
The measurement cycle is repeated $100$ times.

From measurements in co-circular and cross-circular polarizations of the PL detection, the $\rho_c$ dynamics is evaluated [see Fig.~\ref{fig:FastRelaxation:a}].
For further analysis, we use the value of the the electron-spin-polarization degree immediately after the dark interval, $\rho_\mathrm{dark} = \rho_c(t=0)$. 
The Overhauser field immediately after the dark interval can be evaluated from the PL polarization degree~\cite{KoturPRB18},
\begin{equation}
B_N = B_{1/2} \sqrt{\frac{\rho_0 - \rho_\mathrm{dark}}{\rho_\mathrm{dark}-\rho_0\sin^2\theta}} - \sqrt{B_x^2 + B_z^2},
\label{eq:nuc_field}
\end{equation}
where $\rho_0$ and $B_{1/2}$ are obtained in the absence of nuclear spin polarization [see Fig. \ref{fig:PLHanle:b}], and $\theta$ is the angle between the external magnetic field and $x$~axis.

An example of the evolution of $B_N$ calculated using Eq.~\eqref{eq:nuc_field} for different $t_\mathrm{dark}$ is shown in Fig.~\ref{fig:FastRelaxation:b}.
Fitting this value by an exponential function gives the spin relaxation time in the dark due to interaction with electrons, $T_{ND} = 200 \pm 50$~ms.
This time is shorter than $T_\mathrm{bulk}$ shown in Fig.~\ref{fig:Demagnetization:b} by three orders of magnitude and comparable with $T_\mathrm{repump}$ measured at step like polarization modulation. We therefore conclude that the measured times $T_\mathrm{repump}$ and $ T_{ND}$ characterize the relaxation of nuclei under the donor orbit, i.e., they provide an experimental estimate of the hyperfine relaxation time $T_{1e}$ under pumping and in the dark, respectively.
Interpreting $T_\mathrm{repump}$ as characterizing the hyperfine relaxation of nuclei inside the donor orbit in presence of optical excitation, and comparing with the results of Paget \cite{PagetPRB82}, we come to the estimate of the correlation time of the donor-bound electron $\tau_c \approx 200\text{--}300$~ps.
These values agree well both with the data of Ref.~\cite{DzhioevPRB02} and Ref.~\cite{SokolovPRB17}, where $\tau_c$ was determined from polarization recovery under optical pumping in longitudinal magnetic fields.

In contrast, the time $T_{ND}$ at low temperature is, roughly, two times shorter and corresponds to the electron correlation time of about $500$~ps.
This difference is not large as compared to that between the spin-relaxation times of donor-associated and bulk nuclei, but it is indeed noticeable. We attribute it to the shortening of $\tau_c$ by the exchange interaction with itinerant photoexcited electrons~\cite{PagetPRB81, KavokinSST08}.
As discussed in the Introduction, this mechanism was invoked in Ref.~\cite{PagetPRB82} to explain the short electron correlation time found in high-purity GaAs. 
Our result, therefore, further confirms the model of nuclear spin relaxation proposed in Ref.~\cite{PagetPRB82} and expands it to moderately $n$-doped semiconductors.

Another effect of pumping on the nuclear spin relaxation, discussed in Ref.~\cite{PagetAmand2008}, is due to the modulation of the quadrupole splitting of nuclear spins by electric fields induced by charge fluctuations at the donor center.
Such fluctuations may result from recombination of donor-bound electrons with photoexcited holes and subsequent trapping of photoexcited electrons at the donor center that has been emptied by recombination.
Obviously, if this mechanism was effective, the nuclear relaxation under pumping would have been faster than in the dark.
This prediction is opposite to the behavior observed in our experiment.

We must, therefore, conclude that, within the range of pumping levels we use, donor recharging induced by the pump is not frequent enough to induce considerable nuclear spin relaxation.
This mechanism might, however, explain the observed shortening of $T_{ND}$ with temperature increase [see inset to Fig.~\ref{fig:FastRelaxation:b}].
Indeed, at elevated temperatures $T > 4$~K the donor filling factor decreases due to thermal activation of electrons to the conduction band, giving rise to charge fluctuations.
However, the experimental data we have obtained so far are insufficient to conclude whether the observed effect is due to this specific mechanism; we leave this question open for future research.

\section{\label{sec:conclusion}Conclusions}

To conclude, the nuclear spin dynamics has been experimentally studied in $n$-GaAs with donor concentration $n_D = 4 \times 10^{15}$~cm$^{-3}$ by measuring the Hanle effect with millisecond time resolution under time-varying optical pumping.
Two time scales of nuclear buildup and relaxation have been observed.
One of them is slow (on the order of hundreds of seconds) relaxation of bulk nuclei via spin diffusion from (buildup) or to (relaxation) donor centers.
The measured characteristic time $T_\mathrm{bulk} \approx 10^2$~s is typical for the dielectric phase of $n$-GaAs with the studied donor concentration~\cite{VladimirovaPRB17}.

At the same time, a faster dynamics is observed on the timescale from $200$ to $425$~ms, which is attributed to the spin dynamics of nuclei inside the donor orbit.
We also find that the subsecond relaxation time of nuclear spin near donors is faster in the dark than in the presence of optical pumping.
We attribute this effect to exchange scattering of photoexcited itinerant electrons at donors, which shortens the spin correlation time of donor-bound electrons and slows down the hyperfine relaxation.

\begin{acknowledgements}
We acknowledge the financial support of the Deutsche Forschungsgemeinschaft in the frame of the International Collaborative Research Center TRR 160 (Project No. A6), the Russian Foundation for Basic Research (Grant No.~19-52-12043), and Saint-Petersburg State University Research Grant No. ID 28874264 (11.34.2.2012).
\end{acknowledgements}

\end{document}